\documentclass[11pt]{article}
\usepackage{epsfig}
\usepackage{times}
\usepackage{amsfonts}
\usepackage{amsmath}
\usepackage{graphicx}
\usepackage{url}
\usepackage{amssymb}
\usepackage[text={6in,8.5in},centering]{geometry}

\begin{document}
\title{Image recognition with an adiabatic quantum computer\\ I. Mapping to quadratic unconstrained binary optimization}
\author{Hartmut Neven \\
{\it Google, 604 Arizona Blvd. Santa Monica, CA 90401} \\
\\
Geordie Rose and William G. Macready \\
{\it D-Wave Systems Inc., 100 - 4401 Still Creek Drive, Burnaby, BC, Canada V5C 6G9}}

\date{\today}
\maketitle

\begin{abstract}
Many artificial intelligence (AI) problems naturally map to NP-hard optimization problems. This has the interesting consequence that enabling human-level capability in machines often requires systems that can handle formally intractable problems. This issue can sometimes (but possibly not always) be resolved by building special-purpose heuristic algorithms, tailored to the problem in question. Because of the continued difficulties in automating certain tasks that are natural for humans, there remains a strong motivation for AI researchers to investigate and apply new algorithms and techniques to hard AI problems. Recently a novel class of relevant algorithms that require quantum mechanical hardware have been proposed. These algorithms, referred to as quantum adiabatic algorithms, represent a new approach to designing both complete and heuristic solvers for NP-hard optimization problems. In this work we describe how to formulate image recognition, which is a canonical NP-hard AI problem, as a Quadratic Unconstrained Binary Optimization (QUBO) problem. The QUBO format corresponds to the input format required for D-Wave superconducting adiabatic quantum computing (AQC) processors.

\end{abstract}

\section{Introduction}

Humans currently have substantial performance advantages over machines in several areas, including object recognition, knowledge representation, reasoning, learning and natural language processing \cite{Norvig2004}. Intruigingly, most of the hard problems arising in these areas can naturally be cast as NP-hard optimization problems, with the majority reducible to pattern matching problems such as maximum common subgraph \cite{Smith1999, DBLP:conf/gbrpr/2007,Bunke2000, Bunke2008, Sinha2002}. The formal intractability of most problems associated with human intelligence is at the heart of the continued difficulties AI researchers face in mimicking or surpassing human capabilities in these areas.

It may seem surprising that capabilities that we take for granted and perform quite easily could be computationally intractable. However it is important to remember that this intractability does not preclude efficient generation of approximate solutions. In practice, exact solutions to optimization problems arising in AI are not required. Generally there is a graceful degradation of performance as a solution moves away from global optimality. Because of this behavior the ideal computational approach is to use specialized heuristic algorithms to attack these problems \cite{Simon2000}. It is interesting to note that human brains are thought to contain structures specialized for pattern matching (`wetware heuristics') that are used to support a variety of capabilities for which humans still hold a performance advantage over machines, and that these structures have been used as inspirations for development of successful heuristic algorithms \cite{Sinha2002, Mountcastle1997, maclennan1991}.

In this article we focus on the quintessential pattern recognition problem of deciding whether two images contain the same object. This is a typical example of a capability in which humans outperform modern computing systems and can be thought of as an NP-hard optimization problem. We begin to explore whether quantum adiabatic algorithms \cite{Farhi2000, Childs2000, Aeppli1999, Santoro2002} can be employed to obtain better solutions to this problem than can be achieved with classical optimization algorithms. The first step in this exploration is to map image recognition into the particular input format required for running quantum adiabatic algorithms on D-Wave superconducting AQC processors.

\begin{figure}
\begin{center}
\includegraphics[width=6in,angle=0,bb=0 0 435 160]{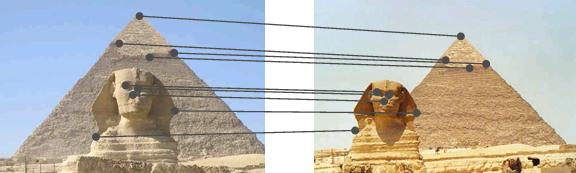}
\caption{Object recognition by image matching proceeds by pairing points in two images that correspond to the same structure in the outside world. In the algorithms considered here, both feature similarity and geometric consistency are considered in determining to what extent two images are similar.}
\end{center}
\label{fig1}
\end{figure}

\section{Image matching}

A popular method to determine whether two images contain the same object is {\it image matching}. Image matching in its simplest form attempts to find pairs of image features from two images that correspond to the same physical structure. An image feature is a vector that describes the neighborhood of a given image location. In order to find corresponding features two factors are typically considered: {\it feature similarity}, as for instance determined by the scalar product between feature vectors, and {\it geometric consistency}. The latter is best defined when looking at rigid objects. In this case the feature displacements are not random but exhibit correlations brought about by a change in viewpoint. For instance, if the camera moves to the left we observe translations of the feature locations in the image to the right. If the object is deformable or articulate then the feature displacements are not solely determined by the camera viewpoint anymore but one can still expect that neighboring features tend to move in a similar way.  Thus image matching can be cast as an optimization problem in which one attempts to minimize an objective function that consists of two terms. The first term penalizes mismatches between features drawn from image one and placed at corresponding locations in image two. The second term enforces spatial consistency between neighboring matches by measuring the divergence between them. It has been shown that this constitutes an NP-hard optimization problem \cite{Felzenzwalb2005}.

\begin{figure}
\begin{center}
\includegraphics[width=6in,angle=0,bb=0 0 1300 450]{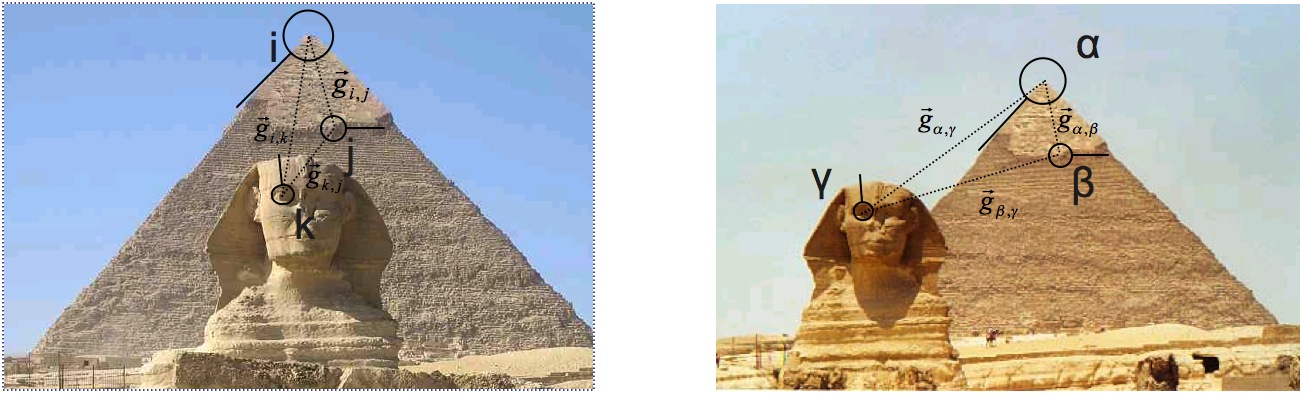}
\caption{Representation of images as labeled graphs. Shown are three exemplary interest points for each image. The number of interest points detected is content dependent but is on the order of several hundred for 640x480 resolution images with content as shown. Each interest point is assigned a position, scale, and orientation \cite{Lowe1999}. In the figure the scale is indicated by a circle and the orientation by a pointer. This information can be used to characterize the relative pose and position of two interest points denoted by the vectors $\vec{g}$ next to the dotted lines.}
\end{center}
\label{fig2}
\end{figure}

\section{Mapping Image Matching to a Quadratic Optimization Problem}

D-Wave AQC processors take as input problems of the form
\begin{equation}
\vec{x}_{opt}=\arg \min \left\{ \sum_{i \leq j=1}^{N} Q_{ij} x_i x_j \right\} \; , \; x_i \in \{0,1\}
\label{eq1}
\end{equation}
which are typically referred to as Quadratic Unconstrained Binary Optimization (QUBO) problems. Physicists will recognize this objective as being closely related to the Ising energy function. In order to use D-Wave hardware to run quantum adiabatic algorithms, the problem of interest must first be converted to this format. In this section we will describe how to perform this conversion for image matching.

\subsection{Representing an image as a labeled graph}

As a first step towards casting image matching problems as QUBO minimization, it is convenient to reduce the amount of data and to focus on features that are sufficiently unique and robust under small image transformations that finding a correspondence is well defined. This reduction to salient image structures is accomplished with an interest point operator. Many versions of interest point operators have been described in the literature \cite{Mikolajczyk2004, Mikolajczyk2005}. In our implementation we use a Laplacian of Gaussians operator. Each interest point $i$ is described by a normalized feature vector  $\vec{f}_i$ called a {\it local descriptor}. We employ a new local descriptor called CONGAS that is based on Gabor wavelets of varying scale and orientation drawn from a space variant grid around an interest point \cite{Buddemeier2008}\cite{Wiskott1997}. Each pair of interest points is associated with a quantity $\vec{g}_{ij}$ that describes the geometric relationship between the interest points. $\vec{g}_{ij}$ measures the translation between interest points as well as the difference in local scale and orientation. The quantity $\vec{g}_{ij}$  is normalized for global translation, rotation and scaling. Using these concepts, any image can be represented by a labeled graph $G$ (see Fig. 2).

\subsection{Generating a conflict graph $G_C$ from two image graphs $G_1$ and $G_2$}

Images $I_1$ and $I_2$ give rise to collections of interest points. We associate the $M$ interest points of $I_1$ with vertices in a labeled graph $G_1$ having $M$ nodes. The $i^{th}$ vertex in $G_1$ is labeled with the feature vector $\vec{f}_i$. Edges in $G_1$ between feature points $i$ and $j$ are labeled with $\vec{g}_{ij}$. Edges characterize the geometric relationships between feature vectors. Similarly, image $I_2$ is represented as a labeled graph $G_2$ over $N$ features (vertices).  In this representation the similarity of two images is specified by the similarity of the two labeled graphs.

For pairs of interest points drawn from each image ($i \in G_1, \alpha \in G_2$) we calculate the similarity of the associated feature vectors $d(i,\alpha)=d_{feat}(\vec{f}_i,\vec{f}_{\alpha})$. Note that when using subscripts we will use the convention that Latin subscripts refer to $G_1$ and Greek subscripts to $G_2$. $d(i,\alpha)$ is a measure of the similarity of the interest points $i$ and $\alpha$. A common choice to measure the distance between two feature vectors, which we assume to be normalized such that $|d(i,\alpha)| \leq 1$, is the scalar product. Of all of the potential matches between interest points in $G_1$ and $G_2$, some will be excellent (corresponding to $d(i,\alpha)$ close to one) and some will be poor. As we do not want to keep poor matches, we introduce a point-wise inclusion threshold $T_{feat}$. We define a pair of points $(i,\alpha)$ to be a potential match if  $d(i,\alpha)> T_{feat}$. Increasing $T_{feat}$ decreases the number of potential matches i.e. it increases the standard for what constitutes an acceptable point-wise match.

\subsection{Generation of vertices in the conflict graph $G_C$}

We generate a conflict graph $G_C$ from $G_1$ and $G_2$ to measure the similarity of the two graphical representations of the images. Starting from the largest $d(i,\alpha)$ we add a vertex $V_{i \alpha}$ to $G_C$ until either all potential matches have been included or the number of vertices reaches a hardware dependent limit $L$. Vertex  $V_{i \alpha}$ corresponds to an association of feature   $i$ in $G_1$ with feature $\alpha$ in $G_2$.

\subsection{Generation of edges in $G_C$}

Edges $(i,\alpha; j,\beta)$ in $G_C$ encode geometric consistency between feature vectors  $\vec{f}_i$ and  $\vec{f}_j$ in $G_1$ and feature vectors $\vec{f}_{\alpha}$ and $\vec{f}_{\beta}$  in $G_2$ . For all vertex pairs $(V_{i \alpha} , V_{j \beta})$  in $G_C$ where $i \neq j$ and $\alpha \neq \beta$ we calculate $d(i,\alpha,j,\beta)=d_{geom}(\vec{g}_{ij}, \vec{g}_{\alpha \beta})$. This quantity we call the {\it geometric consistency} of the two pairs of interest points, and is normalized such that $|d(i,\alpha, j,\beta) | \leq 1$. It measures the geometric compatibility of the match pairs $(i,\alpha)$ and $(j,\beta)$ as the residual differences in local displacement, scale and rotation assignment of the associated interest points after changes due to global translation, rotation and scaling have been normalized out.  A pair $(i, j)$ and $(\alpha, \beta)$ are not allowed to match if they are in geometric conflict i.e. if the residual differences are too large. This requirement is enforced by choosing a threshold $T_{geom}$ whereby if $d(i,\alpha,j,\beta) < T_{geom}$ the pair $(i, j)$ and $(\alpha, \beta)$ are considered to be in geometric conflict. Thus, the prescription for edge drawing in $G_C$ is as follows:
\begin{itemize}
\item For all pairs of vertices $(V_{i \alpha} , V_{j \beta})$  in $G_C$, draw an edge between the vertex pair $(V_{i \alpha} , V_{j \beta})$ if $i = j$ or $\alpha = \beta$
\item For all pairs of vertices $(V_{i \alpha} , V_{j \beta})$  in $G_C$, draw an edge between the vertex pair  $(V_{i \alpha} , V_{j \beta})$ if $i \neq j$ and $\alpha \neq \beta$ and  $(V_{i \alpha} , V_{j \beta})$ are in geometric conflict (ie.  $d(i,\alpha,j,\beta) < T_{geom}$)
\end{itemize}
In either case the presence of an edge records a conflict in associating $i$ with $\alpha$ and $j$ with $\beta$. Note that the first condition ensures that two features which are separate in one image are not be mapped onto a single feature in a second image.

\begin{figure}
\begin{center}
\includegraphics[width=6in,angle=0,bb=0 0 880 400]{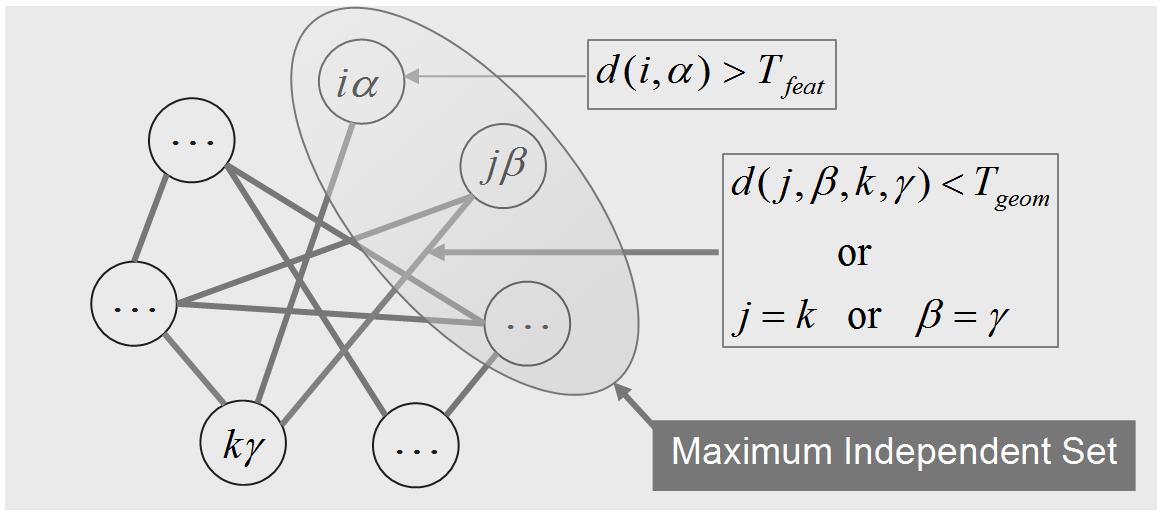}
\caption{Image matching via determining the maximum independent set (MIS) of a conflict graph. The MIS problem is a special case of the more general QUBO problem class. In the MIS formulation each vertex $V_{i \alpha}$ in the graph represents a match candidate for which the feature similarity exceeded a threshold $T_{feat}$. An edge is placed between any two matches that are geometrically inconsistent. The MIS represents the largest set of matches that are geometrically consistent.}
\end{center}
\label{fig3}
\end{figure}

\subsection{Structure of GC}

The graph generated by the preceding prescription we call a {\it conflict graph}. It has by construction at most $L$ vertices, and has arbitrary connectivity with at most $L(L-1)/2$ edges. The maximum independent set (MIS) of the conflict graph is equivalent to the maximum common subgraph of unlabeled graphs $G_1$ and $G_2$. The MIS provides both a similarity measure (the larger the MIS, the greater the region of mutual overlap) and the largest conflict-free mapping of features in $G_1$ to features in $G_2$. For example if the MIS of $G_C$ is the set  $\left\{ V_{i \alpha} , V_{j \beta}, V_{k \gamma} \right\}$ then the point-wise matches between the two graphs are $i \leftrightarrow \alpha$, $j \leftrightarrow \beta$, and $k \leftrightarrow \gamma$. Finding the MIS for the conflict graph $G_C$ can be cast as a QUBO by setting $Q_{i\alpha,i\alpha}=-1$ for all vertices and $Q_{i\alpha,j\beta}=L$ whenever there is an edge between $(i,\alpha)$ and  $(j,\beta)$. The minimum energy configuration enforces $x_{i \alpha}=1$  if and only if $V_{i \alpha} \in MIS$ and $x_{i \alpha}=0$ otherwise. More elaborate objectives defining the correspondence between images can be chosen. Those would not necessarily constitute an MIS problem but could still be formulated as a QUBO.

\section{Summary and Next Steps}

This article presents a description of how to map image recognition problems into the input format required for using D-Wave superconducting AQC processors. This represents a first step in determining to what extent quantum adiabatic algorithms may be useful as components of novel solvers for the NP-hard optimization problems underlying image recognition.

Quantum adiabatic algorithms are to date largely theoretical and verification of their utility as components of either complete or heuristic solvers, and their actual performance in either regime, awaits experimental verification \cite{Lloyd2008}. The next step in our reporting of this work will be a detailed description of the results of solving QUBO instances generated by image matching problems using D-Wave superconducting AQC hardware.

\section*{Acknowledgments}
We would like to thank Ulrich Buddemeier, Alessandro Bissacco, Hartwig Adam and Jinjun Xu for their help with the object recognition routines.

\bibliography{scibib11}

\newcommand{\etalchar}[1]{$^{#1}$}
\begin{thebibliography}{WFKvdM97}

\bibitem[BBTA99]{Aeppli1999}
J.~Brooke, D.~Bitko, T.F.Rosenbaum, and G.~Aeppli.
\newblock Quantum annealing of a disordered magnet.
\newblock {\em Science}, 284:779--781, 30 April 1999.

\bibitem[BDK{\etalchar{+}}08]{Bunke2008}
Horst Bunke, Peter Dickinson, Miro Kraetzl, Michel Neuhaus, and Marc Stettler.
\newblock {\em Matching of Hypergraphs--Algorithms, Applications, and
  Experiments}, volume~91 of {\em Studies in Computational Intelligence}.
\newblock Springer Berlin/Heidelberg, 2008.

\bibitem[BN08]{Buddemeier2008}
Ulrich Buddemaier and Hartmut Neven.
\newblock System and method for descriptor vector computation.
\newblock {\em US Patent Application}, 2008.

\bibitem[Bun00]{Bunke2000}
Horst Bunke.
\newblock Graph matching: Theoretical foundations, algorithms, and
  applications.
\newblock {\em Proc. Vision Interface}, pages 82--88, 2000.

\bibitem[CFGG00]{Childs2000}
A.M. Childs, E.~Farhi, J.~Goldstone, and S.~Gutmann.
\newblock Finding cliques by quantum adiabatic evolution.
\newblock 2000.
\newblock preprint quant-ph/0012104.

\bibitem[EFS00]{Farhi2000}
Sam~Gutmann Edward~Farhi, Jeffrey~Goldstone and Michael Sipser.
\newblock Quantum computation by adiabatic evolution.
\newblock 2000.
\newblock preprint quant-ph/0001106v1.

\bibitem[EV07]{DBLP:conf/gbrpr/2007}
Francisco Escolano and Mario Vento, editors.
\newblock {\em Graph-Based Representations in Pattern Recognition, 6th
  IAPR-TC-15 International Workshop, GbRPR 2007, Alicante, Spain, June 11-13,
  2007, Proceedings}, volume 4538 of {\em Lecture Notes in Computer Science}.
  Springer, 2007.

\bibitem[FH05]{Felzenzwalb2005}
Felzenzwalb and Huttenlocher.
\newblock Pictorial structures for object recognition.
\newblock {\em Intl. Journal of Computer Vision}, 61(1):55--79, 2005.

\bibitem[Llo08]{Lloyd2008}
Seth Lloyd.
\newblock Quantum information matters.
\newblock {\em Science}, 319:1209--1211, 29 February 2008.

\bibitem[Low99]{Lowe1999}
David~G. Lowe.
\newblock Object recognition from local scale-invariant features.
\newblock {\em International Conference on Computer Vision}, pages 1150--1157,
  September 1999.

\bibitem[Mac91]{maclennan1991}
Bruce MacLennan.
\newblock Gabor representations of spatiotemporal visual images.
\newblock Technical Report UT-CS-91-144, 1991.

\bibitem[Mou97]{Mountcastle1997}
VB~Mountcastle.
\newblock The columnar organization of the neocortex.
\newblock {\em Brain}, 120:701--722, 1997.

\bibitem[MS04]{Mikolajczyk2004}
K.~Mikolajczyk and C.~Schmid.
\newblock Scale and affine invariant interest point detectors.
\newblock {\em International Journal of Computer Vision}, 60(1):63--86, 2004.

\bibitem[MTS{\etalchar{+}}05]{Mikolajczyk2005}
K.~Mikolajczyk, T.~Tuytelaars, C.~Schmid, A.~Zisserman, J.~Mats,
  F.~Schaffalitzky, T.~Kadir, and L.~Van Gool.
\newblock A comparison of affine region detectors.
\newblock {\em International Journal of Computer Vision}, 65, 2005.

\bibitem[RN03]{Norvig2004}
Stuart Russell and Peter Norvig.
\newblock {\em Artificial Intelligence: A Modern Approach}.
\newblock Prentice-Hall, Inc., 2nd edition, 2003.

\bibitem[Sim95]{Simon2000}
Herbert~A. Simon.
\newblock Artificial intelligence: an empirical science.
\newblock {\em Artificial Intelligence}, 77:95--127, August 1995.

\bibitem[Sin02]{Sinha2002}
Pawan Sinha.
\newblock Recognizing complex patterns.
\newblock {\em Nature Neuroscience}, 5:1093--1097, 2002.

\bibitem[Smi99]{Smith1999}
Kate~A. Smith.
\newblock Neural networks for combinatorial optimization: A review of more than
  a decade of research.
\newblock {\em INFORMS Journal on Computing}, 11(1):15--34, Winter 1999.

\bibitem[SMTC02]{Santoro2002}
Guiseppe~E. Santoro, Roman Martonak, Erio Tosatti, and Roberto Car.
\newblock Theory of quantum annealing of an ising glass.
\newblock {\em Science}, 295:2427--2430, 29 March 2002.

\bibitem[WFKvdM97]{Wiskott1997}
L.~Wiskott, J.-M. Fellous, N.~Kruger, and C.~von~der Malsburg.
\newblock Face recognition by elastic bunch graph matching.
\newblock In {\em ICIP '97: Proceedings of the 1997 International Conference on
  Image Processing (ICIP '97) 3-Volume Set-Volume 1}, page 129, Washington, DC,
  USA, 1997. IEEE Computer Society.

\end{thebibliography}
\bibliographystyle{alpha}

\end{document}